\documentclass[reprint,superscriptaddress,aps,twocolumn,showpacs]{revtex4-1}
\usepackage{amssymb}
\usepackage{amsmath}
\usepackage{graphicx}
\graphicspath{{figures/}} 
\usepackage{hyperref}
\usepackage{braket,mleftright}
\usepackage{xcolor}
\usepackage{bm}

\allowdisplaybreaks

\begin{document}
	
\title{Study of $N(1440)$ structure via $\gamma^*p\to N(1440)$ transition}
	
\author{A. Kaewsnod}
\email[]{a.kaewsnod@gmail.com}
\author{K. Xu}
\author{Z. Zhao}
\affiliation{School of Physics and Center of Excellence in High Energy Physics and Astrophysics, Suranaree University of Technology, Nakhon Ratchasima 30000, Thailand}
\author{X. Y. Liu}
\affiliation{School of Physics and Center of Excellence in High Energy Physics and Astrophysics, Suranaree University of Technology, Nakhon Ratchasima 30000, Thailand}
\affiliation{School of Physics, Liaoning University, Shenyang 110036, China}
\author{S. Srisuphaphon}
\affiliation{Department of Physics, Faculty of Science, Burapha University, Chonburi 20131, Thailand}
\author{A. Limphirat}
\author{Y. Yan}
\email[]{yupeng@g.sut.ac.th}
\affiliation{School of Physics and Center of Excellence in High Energy Physics and Astrophysics, Suranaree University of Technology, Nakhon Ratchasima 30000, Thailand}
	
\date{\today}
	
\begin{abstract}
We study the photoproduction of the $N(1440)$ resonance in $\gamma^*p\to N^*$ process in quark models, where the $N(1440)$ takes different wave functions: first radial excitation of the nucleon imported from low-lying baryon mass spectrum calculations, a general radial excitation of the nucleon, and a $q^3$ state with positive parity.
The comparison between the theoretical results and experimental data on the helicity amplitudes $A_{1/2}$ and $S_{1/2}$ and the analysis of the spatial wave function of the $N(1440)$ resonance reveal that the $N(1440)$ resonance is mainly the $q^3$ first radial excitation.

\end{abstract}
	
\maketitle
	
\section{\label{sec:Int} Introduction}
In recent years, the experiments on electro- and photoproductions of particles have accumulated a large amount of data of helicity amplitudes over the four-momentum transfer $Q^2$ of virtual photons \cite{Zyla:2020zbs,PhysRevC.80.055203,PhysRevC.86.035203,PhysRevC.93.025206}.
The most complete experimental data has been obtained for the Roper resonance, $N(1440)$, up to 4.5 GeV$^2$ \cite{Tiator:2011pw,Tiator:2009mt}.
Theoretical studies have been carried out to reveal the properties and structures of nucleon resonance states by analyzing the transverse transition amplitudes $A_{1/2}$ and $A_{3/2}$ and longitudinal transition amplitude $S_{1/2}$ over $Q^2$ in a large range \cite{PhysRevC.76.025212,PhysRevC.78.045209,Capstick_2007,PhysRevD.81.074020,Golli:2009uk,RevModPhys.91.011003,PhysRevC.74.015202,PhysRevD.84.014004,Cano:1998wz,PhysRevLett.119.022001}.
However, the electro- and photoproduction data of the Roper resonance have not been well described by the conventional quark models \cite{PhysRevC.76.025212,PhysRevC.78.045209,Capstick_2007,PhysRevD.81.074020,Golli:2009uk} as well as other approaches.
The nature of the Roper resonance is still an open question, which is also implied by its large decay width \cite{Sarantsev:2007aa} as well as the mass ordering with the lowest negative-parity baryon resonance states $N(1520)$ and $N(1535)$ \cite{Capstick:2000qj}. Detailed discussion about the Roper resonance may be found in a good review paper \cite{RevModPhys.91.011003}.
Among a large number of theoretical works, the light-front relativistic quark model \cite{PhysRevC.76.025212,PhysRevC.78.045209,Capstick_2007} and the covariant spectator quark model \cite{PhysRevD.81.074020}, assuming that the $N(1440)$ is the first radial excitation of the $q^3$ ground state, can give the right sign of the transverse $\gamma^*p\to N(1440)$ amplitudes and reproduce the experimental data of the transverse amplitudes at high $Q^2$, but fail to describe the data of the $N(1440)$ helicity amplitude at low $Q^2$.
	
Our previous works \cite{PhysRevC.100.065207,PhysRevD.101.076025}, where the mass ordering problem of $N(1440)$, $N(1520)$, and $N(1535)$ is tackled by including ground state light pentaquark components in the negative-parity nucleon resonances, point out that the $N(1440)$ might be mainly the first radial excitation of the nucleon.
In this work, we push further by studying the photoproduction transitions $\gamma^*p\to N^*$ in the three-quark picture. 	
The paper is organized as follows.
In Sec.~\ref{sec:Form}, we briefly present the formalism of the proton electric form factor, the longitudinal and transverse helicity amplitudes of the $N(1440)$ resonance in the constituent quark model.
The proton $G_E$ and the $A_{1/2}$ and $S_{1/2}$ of the $N(1440)$ resonance are evaluated in Sec. \ref{sec:LLB Result}, where the wave functions of the proton as the ground state of three quarks and the Roper resonance as the first radial excitation of the nucleon are taken from the study of the low-lying baryon mass spectra.
In Sec. \ref{sec:RCL0 Result}, we study the $A_{1/2}$ and $S_{1/2}$ of the $N(1440)$ resonance by assigning the roper resonance as a general radial excitation of the nucleon as well as just a $q^3$ state with positive parity.
The quark distribution of the $N(1440)$ is derived in Sec. \ref{sec:RCL0 Result} from the results in Sec. \ref{sec:LLB Result} and Sec. \ref{sec:RCL0 Result}.
A summary is given in Sec.~\ref{sec:Sum}.
	
\section{\label{sec:Form} Formalism of helicity amplitudes and electromagnetic form factor}
\begin{figure}[t!]
	\centering
	\includegraphics[width=0.4\textwidth]{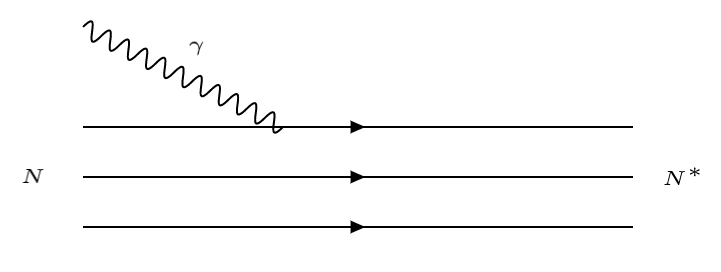}
	\caption{Diagram of photoproduction transition $N\gamma^*\rightarrow N^*$ where both the initial nucleon ($N$) and final nucleon resonance ($N^*$) are in the three-quark picture.}\label{fig:quarkline}
	\end{figure}

We study the $N\gamma^*\rightarrow N^*$ electromagnetic transition in the constituent quark model.
The photoproduction diagram is shown in Fig. \ref{fig:quarkline}.
Both the nucleon and resonance states are composed of three quarks.
	
The proton electric form factor $G_E$, the time component of the electromagnetic current, is derived in the Breit frame.
The momenta of the initial-state nucleon, final-state nucleon, and photon are respectively defined as $P_i=(E_N,0,0,-|\bm k/2|)$, $P_f=(E_N,0,0,|\bm k/2|)$, $k=(0,0,0,|\bm k|)$ in the Breit frame, with $E_N$ and $\bm k$ being the energy and three-momentum of the photon.
The photon energy is $\omega=0$ as $\bm P_i+\bm P_f=0$ in the Breit frame.
The square of the four-momentum transfer is expressed by $Q^2=-k^2=|\bm{k}|^2$.	
The transverse helicity amplitude $A_{1/2}$ and the longitudinal helicity amplitude $S_{1/2}$ are usually defined in the $N^*$ rest frame.
The momentum of the nucleon $P_i$, the momentum of the resonance state $P_f$, and the photon momentum $k=P_f-P_i$ are defined as $P_i=(E_N,0,0,-|\bm k|)$, $P_f=(M_{N^*},0,0,0)$, $k=(\omega,0,0,|\bm k|)$.
One has
\begin{alignat}{3}\label{eq:p photon}
	&E_N&&=\ \frac{M_{N^*}^2+M_N^2+Q^2}{2M_{N^*}},\nonumber\\
	&\omega&&=\ \frac{M_{N^*}^2-M_N^2-Q^2}{2M_{N^*}},\nonumber\\
	&|\bm k|&&=\ \left[ Q^2+\left( \frac{M_{N^*}^2-M_N^2-Q^2}{2M_{N^*}}\right)^2\right]^\frac{1}{2}.
\end{alignat}

The proton electric form factor, the transverse and longitudinal helicity amplitudes of electromagnetic transitions of the $\gamma^*p\to N^*$ take the form,
\begin{alignat}{3}
	&G_E&&=&&\ \Braket{N,S'_z=\frac{1}{2}|q_1'q_2'q_3'}T_B(q_1q_2q_3\rightarrow q_1'q_2'q_3')\nonumber\\
	&&&&&\left. \times \Braket{q_1q_2q_3|N,S_z=\frac{1}{2}}\right|_\text{Breit frame},\nonumber\\
	&A_{1/2}&&=&&\ \dfrac{1}{\sqrt{2K}}\Braket{N^*,S'_z=\frac{1}{2}|q_1'q_2'q_3'}\nonumber\\
	&&&&&\times T(q_1q_2q_3\rightarrow q_1'q_2'q_3')\Braket{q_1q_2q_3|N,S_z=-\frac{1}{2}},\nonumber\\
	&S_{1/2}&&=&&\ \dfrac{1}{\sqrt{2K}}\dfrac{|\bm k|}{Q}\Braket{N^*,S'_z=\frac{1}{2}|q_1'q_2'q_3'}\nonumber\\
	&&&&&\times T(q_1q_2q_3\rightarrow q_1'q_2'q_3')\Braket{q_1q_2q_3|N,S_z=\frac{1}{2}},
	\label{eq:helicity}
\end{alignat}
with
\begin{alignat}{3}
	K=\frac{M_{N^*}^2-M_N^2}{2M_{N^*}}
\end{alignat}
where $K$ is the real-photon momentum in the $N^*$ rest frame, $M_{N^*}$ and $M_N$ are respectively the $N^*$ and $N$ masses, and $\Braket{q_1q_2q_3|N,S_z}$ and $\Braket{q_1'q_2'q_3'|N^*,S'_z}$ are the $N$ and $N^*$ wave functions in the three-quark picture.
$T(q_1q_2q_3\rightarrow q_1'q_2'q_3')$ in Eq. (\ref{eq:helicity}) is the transition amplitude of the process $\gamma q\to q'$ displayed in Fig. \ref{fig:quarkline}, which can be calculated in the standard language of quantum field theory,
\begin{align}\label{eq:Matrix elements}
	T(q_1q_2q_3\rightarrow q_1'q_2'q_3')=\ &e_3\bar u_{s'}(p')\gamma^\mu u_s(p)\epsilon_\mu^\lambda(k)\Braket{q'_1q'_2|q_1q_2}\nonumber\\
	=\ &e_3T_{s's}^\lambda\Braket{q'_1q'_2|q_1q_2},
\end{align}
where $e_3$ and $u_{s(s')}$ are the electric charge and the Dirac spinners of the third quark with $s',\,s$ being the single quark spin projections (spin up $\uparrow$ and spin down $\downarrow$), and $\lambda$ is the helicity of the photon.
In the work, the mass of $u$ and $d$ quarks are taken to be $m=5$ MeV.
The photon polarization vectors $\epsilon_\mu^\lambda(k)$ of the longitudinal ($\lambda=0$) and transverse ($\lambda=1$) helicity amplitudes in the Lorentz gauge are defined as $\epsilon^0_\mu=\frac{1}{Q}(|\bm k|,0,0,\omega)$ and $\epsilon^+_\mu=-\frac{1}{\sqrt{2}}(0,1,i,0)$.
The matrix elements $T_{s's}^\lambda$ are shown in Appendix \ref{app:T helicity}.
In Eq. (\ref{eq:helicity}), one sums over all possible quantum states of the intermediate three quarks and integrates over the momenta of all quarks.
	
For the proton electric form factor $G_E$, the matrix elements of electromagnetic transition in the Breit frame are represented as
\begin{align}\label{eq:TB}
	T_B(q_1q_2q_3\rightarrow q_1'q_2'q_3')=\ &e_3\bar u_{s'}(p')\gamma^\mu u_s(p)\epsilon_\mu(k)\Braket{q'_1q'_2|q_1q_2}\nonumber\\
	=\ &e_3T_{s's}^B\Braket{q'_1q'_2|q_1q_2}.
\end{align}
The polarization vector of the photon in Eq. \ref{eq:TB} is defined as $\epsilon_\mu=(1,0,0,0)$.
The matrix elements $T_{s's}^B$ are shown in Appendix \ref{app:T form factor}.
	
The proton charge form factor is calculated in the $q^3$ picture, and the proton wave function is nailed down by comparing the theoretical results with experimental data.
The helicity amplitudes $A_{1/2}$ and $S_{1/2}$ of the $N(1440)$ resonance will be evaluated in the next sections with different $N(1440)$ wave functions: first radial excitation of the nucleon imported from low-lying baryon mass spectrum calculations, a general radial excitation of the nucleon, and a $q^3$ state with positive parity.
		
\section{\label{sec:LLB Result}Helicity amplitudes with $N(1440)$ wave function from mass spectrum study}
In this section, we evaluate the proton electric form factor and the helicity amplitudes with the spatial wave functions of the proton and the Roper resonance derived in the study of low-lying baryon mass spectra \cite{PhysRevD.101.076025}, where the proton is assumed to be the three-quark ground state and the $N(1440)$ to be the first radial excitation of the nucleon.
There is no free parameter in the calculations.
\begin{figure}[t]
	\centering
	\includegraphics[width=0.4\textwidth]{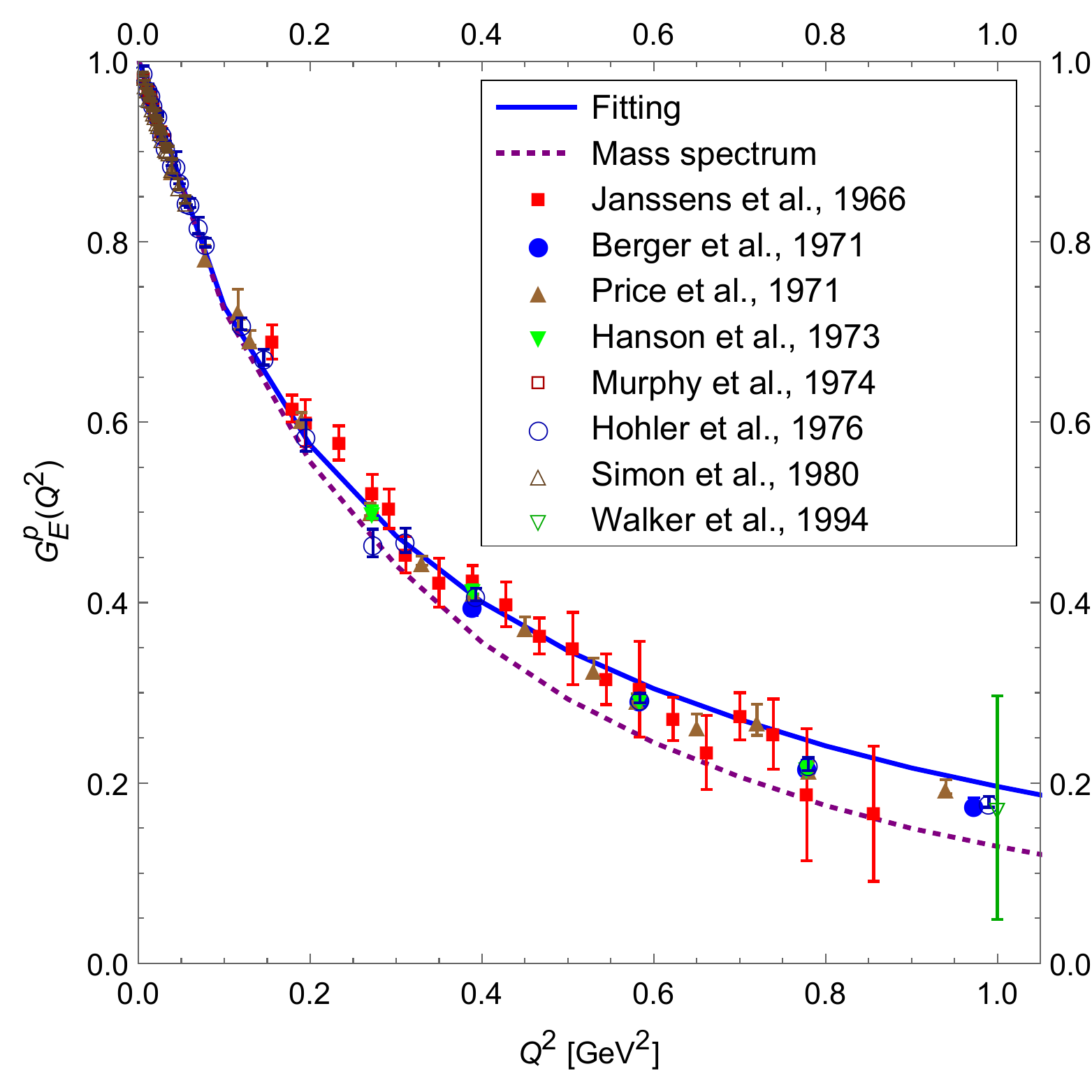}
	\caption[Proton electric form factor]{Proton electric form factor $G_E$ compared to experimental data. The solid curve is the fitting result that leads to the proton spatial wave function to be used in the helicity amplitude calculations, and the short dashed curve is derived with the proton spatial wave function from low-lying baryon mass spectrum calculations. The experimental data are taken from \cite{PhysRev.142.922,Berger:1971kr,Price:1971zk,PhysRevD.8.753,PhysRevC.9.2125,HOHLER1976505,Simon:1980hu,PhysRevD.49.5671}.}
	\label{fig:GE}
\end{figure}
	
The calculated proton electric form factor $G_E$ with the proton spatial wave function imported from the mass spectra fitting is shown as the dotted curve in Fig. \ref{fig:GE} while the theoretical results of transverse and longitudinal helicity amplitudes of the $N(1440)$ resonance are shown as the dotted curve in Figs. \ref{fig:A12} and \ref{fig:S12}, respectively.
It is found in Fig. \ref{fig:GE} that the theoretical results for the proton electric form factor $G_E$ are consistent with the experimental data.
It is a surprise that the proton wave function derived from the baryon mass spectrum calculation leads to perfect results for the electric form factor at lower $Q^2$.
One may believe that meson cloud contributions may account for a considerable part of the proton electric form factor at lower $Q^2$.
However, this work reveals that the three-quark core dominantly contributes to the proton electric form factor, even for very low $Q^2$, and meson cloud contributions may be negligible.

For both the $A_{1/2}$ and $S_{1/2}$, the theoretical results can fairly describe the experimental data at large $Q^2$ region ($1.5 \leq Q^2 \leq 4.5$ GeV$^2$).
This is consistent with the theoretical results in the light-front relativistic quark model \cite{PhysRevC.76.025212} and covariant spectator quark model \cite{PhysRevD.81.074020}.
The theoretical results of the helicity amplitude $A_{1/2}$ are in the right tendency of the $A_{1/2}$ data and can give the right sign at the real photon point though the discrepancy between the theoretical results and the experimental data is obvious.
Considering that the $N(1440)$ wave function employed in the calculation is imported, without any modification, from the baryon mass spectrum calculation, one may have a good belief that the $N(1440)$ is mainly a three-quark state, the first radial excitation of the proton.
The theoretical results of the $S_{1/2}$ at low $Q^2$ region are largely inconsistent with the experimental data, giving a much bigger value at the real photon point, which may indicate that the $N(1440)$ might have some other components rather than the first radial excitation of the nucleon in the three quark picture.
However, it is not necessary to jump to the point that the meson cloud is important among the additional components.
Note that the three quark ground state of the proton derived in the baryon mass spectrum calculations reproduces well the experimental data of the proton charge form factor at small $Q^2$.
		
\section{\label{sec:RCL0 state} Helicity amplitudes with $N(1440)$ as three quark states}
In this section, we calculate the helicity amplitudes of the $N(1440)$ which is allowed to include higher radial excitations beside the first radial excitation of the nucleon as well as just a three quark state of positive parity.
As shown in the previous section, the proton wave function derived from the baryon mass spectrum calculation leads to a reasonable, but not perfect, proton charge form factor.
Here we first extract the proton spatial wave function in the three-quark picture by fitting the theoretical result of the proton electric form factor to the experimental data.
This extracted proton spatial wave function will be employed to study the $N(1440)$.
The proton spatial wave function is extracted from the form factor shown as the solid curve in Fig. \ref{fig:GE}.
	
In this calculation we assume that the $N(1440)$ is in the three-quark configuration, but may contain higher radial excitations of the nucleon or may be just a positive-parity state.
We let the $N(1440)$ spatial wave function be free for fitting the experimental data of the helicity amplitudes of the $N(1440)$ resonance.
We do two calculations: \\
\indent Fitting I: the proton and the $N(1440)$ spatial wave functions are orthogonal each other, $\Braket{\psi^O_{p}|\psi^O_{N(1440)}}=0$ which constrains the $N(1440)$ spatial wave function to be the radial excitations of the nucleon;\\
\indent Fitting II: the proton and the $N(1440)$ spatial wave functions may not be orthogonal each other, which allows the $N(1440)$ spatial wave function composed of the radial excitation of the nucleon as well as others.
\begin{figure}[t]
	\centering
	\includegraphics[width=0.4\textwidth]{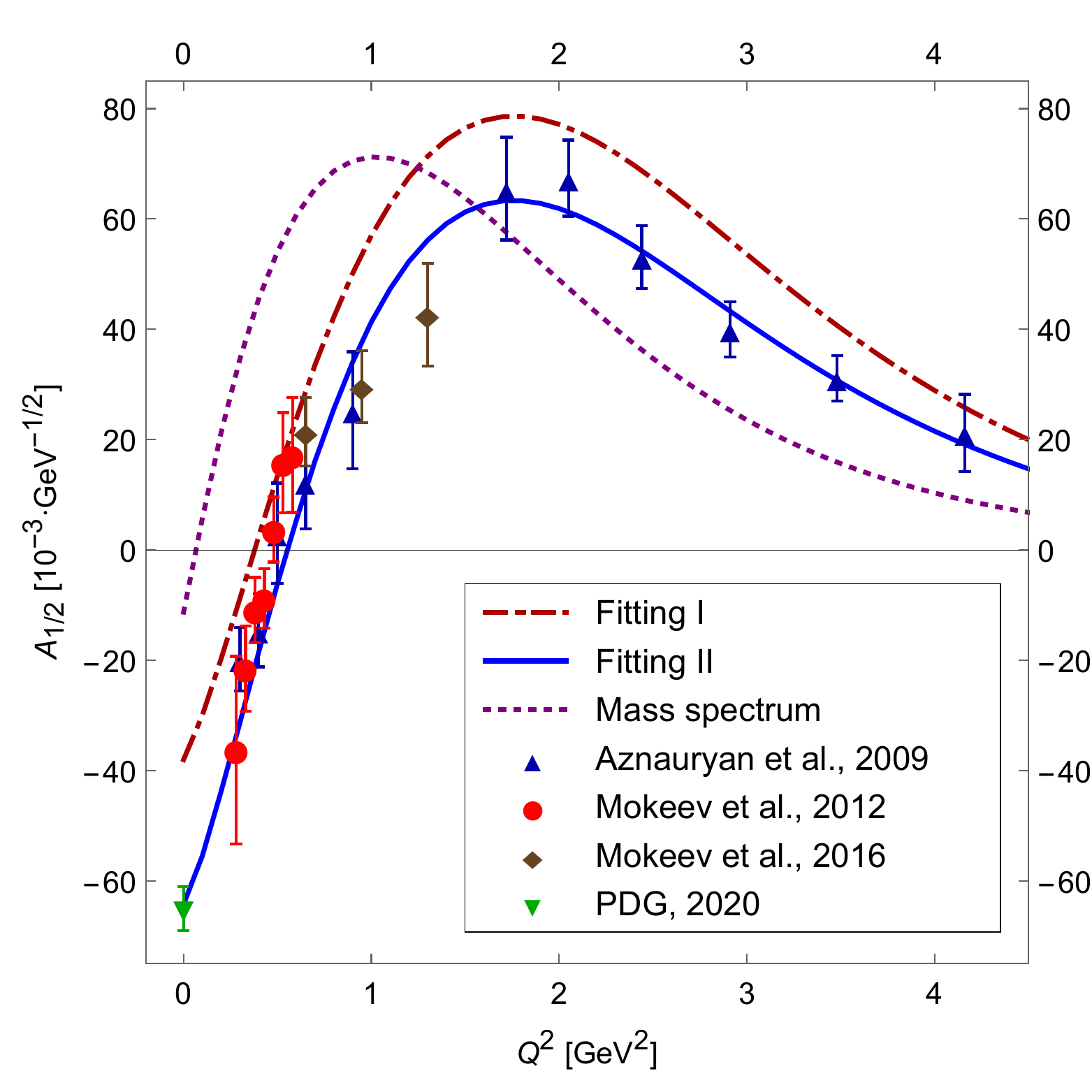}
	\caption[The transverse helicity amplitude of $N(1440)$]{Transverse helicity amplitude $A_{1/2}$ of the $p\gamma^*\rightarrow N(1440)$ transition compared to the measurements. The short dashed, dash-dotted, and solid curves are the results with the $N(1440)$ spatial wave function from the low-lying baryon mass spectrum calculations, in Fitting I, and in Fitting II, respectively. The experimental data are taken from \cite{PhysRevC.80.055203,PhysRevC.86.035203,PhysRevC.93.025206,Zyla:2020zbs}.}
	\label{fig:A12}
\end{figure}

\begin{figure}[t]
	\centering
	\includegraphics[width=0.4\textwidth]{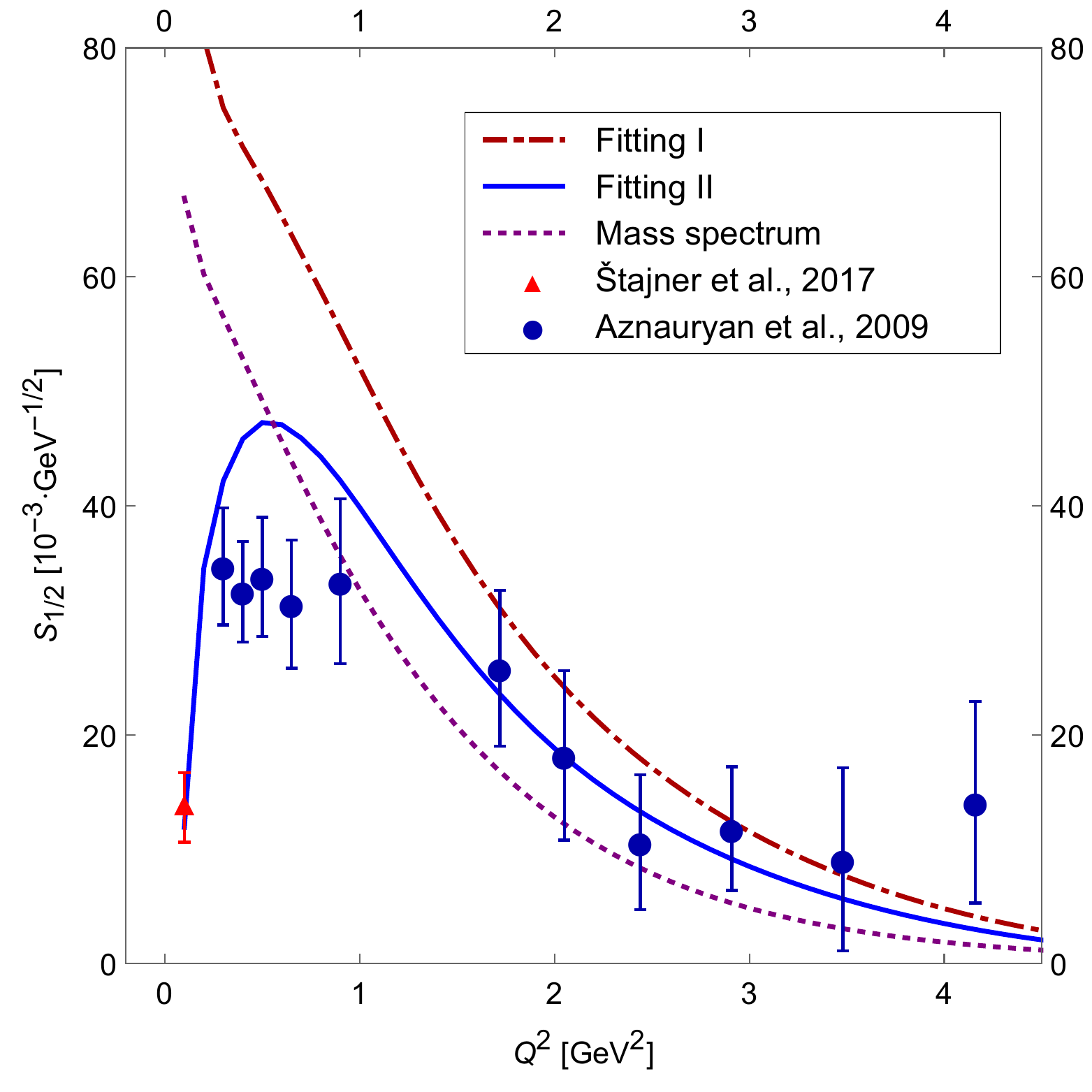}
	\caption[The longitudinal helicity amplitude of $N(1440)$]{Longitudinal helicity amplitude $S_{1/2}$ of $p\gamma^*\to N(1440)$ transition in comparison with experimental data  \cite{PhysRevLett.119.022001,PhysRevC.80.055203}. The short dashed, dash-dotted, and solid curves are the results with the $N(1440)$ spatial wave function from the low-lying baryon mass spectrum calculations, in Fitting I, and in Fitting II, respectively.}
	\label{fig:S12}
\end{figure}

The theoretical results of the helicity amplitudes of the $N(1440)$ resonance are shown in Figs. \ref{fig:A12} and \ref{fig:S12} with the dash-dotted lines for Fitting I and the solid lines for Fitting II.
It is found that the theoretical results of the helicity amplitude $A_{1/2}$ in Fitting I improve more or less the results in Section \ref{sec:LLB Result}, especially for small $Q^2$, but for the amplitude $S_{1/2}$ the results here are rather similar to the ones in Section \ref{sec:LLB Result} without any improvement.
Therefore, one may conclude that the results in Fitting I rule out that the $N(1440)$ may be composed of a sizeable component of higher radial excitations of the nucleon.
	
As shown in Figs. \ref{fig:A12} and \ref{fig:S12}, Fitting II results are in a perfect fit to the amplitude $A_{1/2}$ and a very reasonable fit to the amplitude $S_{1/2}$, particularly to the two values at the real photon point.
The results imply that the $N(1440)$ may consist of components other than radial excitations of three quarks.
	
\section{\label{sec:RCL0 Result} Extraction of quark distribution of $N(1440)$}
\begin{figure}[t]
	\centering
	\includegraphics[width=0.39\textwidth]{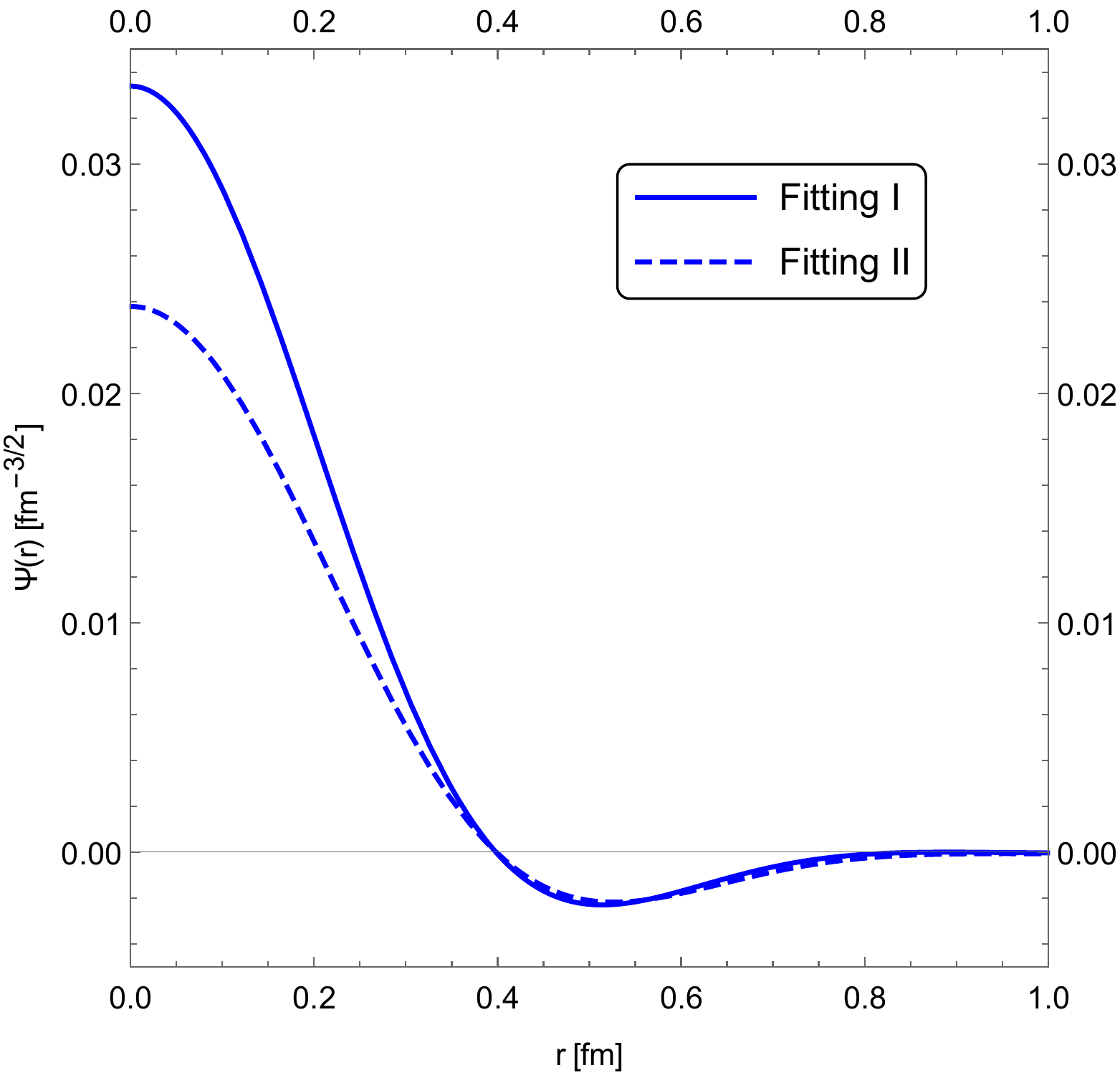}
	\caption[The spatial wave functions of $N(1440)$]{$N(1440)$ spatial wave functions with respect to the relative distance between the quark to the center of mass: solid curve from Fitting I and dashed curve from Fitting II.}
	\label{fig:WF}
\end{figure}

To see the inner structure of the $N(1440)$ closely, we extract the spatial wave functions from the calculations in Section \ref{sec:LLB Result} and \ref{sec:RCL0 state}.
Shown in Fig. \ref{fig:WF} are the $N(1440)$ spatial wave functions with respect to the relative distance between the quark to the center of mass, where the solid and dashed curves are for the $N(1440)$ in Fitting I and Fitting II, respectively.
Considering the full permutation symmetry of the three-quark baryon state, we set
$|\bm{r_\lambda}|=|\bm{r_\rho}|$, where $\bm{r_\rho}=\frac{1}{\sqrt{2}}(\bm{r}_1-\bm{r}_2)$ and $\bm{r_\lambda}=\frac{1}{\sqrt{6}}(\bm{r}_1+\bm{r}_2-2\bm{r}_3)$ to project the three-dimensional spatial wave functions to a 2D plot.
The non-orthogonality between the proton spatial wave function and the $N(1440)$ one derived in Fitting II is $|\braket{\psi^O_{p}|\psi^O_{N(1440)}}|^2=0.03$, indicting that the $N(1440)$ may be composed of components rather than the radial excitations of three quarks.

Fig. \ref{fig:WF} shows that the quark density about the center of mass in Fitting II is lower than the one in Fitting I, and that the $N(1440)$ spatial wave functions are almost the same for larger
distances starting from the node.
The results may tell that it is not the skirt but the very inner part of the $N(1440)$ which accounts for the helicity amplitudes at very small $Q^2$.

\section{\label{sec:Sum} Summary}
In this work, we have studied the proton electric form factor and the helicity amplitudes of the $N(1440)$ resonance via the process $p\gamma^*\to N^*$.
The spatial wave functions of the proton and $N(1440)$ are extracted by fitting the theoretical results to the experimental data of the proton electric form factor and helicity amplitudes.
	
It is found that the proton wave function derived in the baryon mass spectrum study in the three-quark picture leads to an almost perfect proton electric form factor, particularly at low $Q^2$.
One may conclude that the meson cloud contribution to the proton electric form factor is negligible.

The work supports the argument that the $N(1440)$ resonance is mainly the first radial excited state of the nucleon.
However, the Roper may be composed of small components rather than radial excitations of three quarks, such as $D$-wave three quark states, hadronic molecule states, and pentaquarks.
Further studies including these components into the N1440 are under way.
	
\section*{ACKNOWLEDGMENTS}
A.K. and Y.Y. would like to thank for the support of the Thailand Science Research and Innovation (TSRI) for their support through the Royal Golden Jubilee Ph.D. (RGJ-PHD) Program (Grant No. PHD/0242/2558) and A.K. would like to thank for the support of Suranaree University of Technology under SUT-OROG scholarship (contract no. 46/2558).
	
\begin{appendix}
\section{Matrix elements of $\gamma q\to q'$ for helicity amplitudes}\label{app:T helicity}
The matrix elements of the electromagnetic transition $\gamma q\to q'$ for the helicity $\lambda=0,1$ are derived in detail,
\begin{alignat}{5}\label{eq:T}
	&T^0_{\uparrow\uparrow}&&=&&\ \left[ \dfrac{(E'+m)(E+m)}{4E'E}\right]^\frac{1}{2}\left[\dfrac{|\bm k|}{Q}\left( 1+ \dfrac{p_z'p_z+2p_-'p_+}{(E'+m)(E+m)}\right)\right. \nonumber\\
	&&&&&\ \left. -\dfrac{\omega}{Q}\left(  \dfrac{p_z}{E+m}+\dfrac{p_z'}{E'+m}\right) \right], \nonumber\\
	&T^0_{\uparrow\downarrow}&&=&&\ \left[ \dfrac{(E'+m)(E+m)}{4E'E}\right]^\frac{1}{2}\left[\dfrac{|\bm k	|}{Q}\left(  \dfrac{\sqrt{2}(p_z'p_--p_-'p_z)}{(E'+m)(E+m)}\right)\right. \nonumber\\
	&&&&&\ \left.  -\dfrac{\omega}{Q}\left( \dfrac{\sqrt{2}p_-}{E+m}-\dfrac{\sqrt{2}p_-'}{E'+m}\right) \right], \nonumber\\
	&T^0_{\downarrow\uparrow}&&=&&\ \left[ \dfrac{(E'+m)(E+m)}{4E'E}\right]^\frac{1}{2}\left[\dfrac{|\bm k	|}{Q}\left(  \dfrac{\sqrt{2}(-p_z'p_++p_+'p_z)}{(E'+m)(E+m)}\right)\right. \nonumber\\
	&&&&&\ \left.  -\dfrac{\omega}{Q}\left( \dfrac{-\sqrt{2}p_+}{E+m}+\dfrac{\sqrt{2}p_+'}{E'+m})\right)\right], \nonumber\\
	&T^0_{\downarrow\downarrow}&&=&&\ \left[ \dfrac{(E'+m)(E+m)}{4E'E}\right]^\frac{1}{2}\left[\dfrac{|\bm k	|}{Q}\left( 1+\dfrac{p_z'p_z+2p_+'p_-}{(E'+m)(E+m)}\right)\right. \nonumber\\
	&&&&&\ \left.  -\dfrac{\omega}{Q}\left(\dfrac{p_z}{E+m}+\dfrac{p_z'}{E'+m}\right)\right], \nonumber\\
	&T^+_{\uparrow\uparrow}&&=&&\ \left[ \dfrac{(E'+m)(E+m)}{4E'E}\right]^\frac{1}{2}\left[ \dfrac{2p_+}{E+m}\right], \nonumber\\
	&T^+_{\uparrow\downarrow}&&=&&\ \left[ \dfrac{(E'+m)(E+m)}{4E'E}\right]^\frac{1}{2}\left[ -\dfrac{\sqrt{2}p_z}{E+m}+\dfrac{\sqrt{2}p_z'}{E'+m}\right], \nonumber\\
	&T^+_{\downarrow\uparrow}&&=&&\ 0,\nonumber\\
	&T^+_{\downarrow\downarrow}&&=&&\ \left[ \dfrac{(E'+m)(E+m)}{4E'E}\right]^\frac{1}{2}\left[ \dfrac{2p_+'}{E'+m}\right],
\end{alignat}
where $E$ and $E'$ are respectively the energies of the initial and final interacting quarks with the dynamical quark mass of $u$ and $d$ quarks as $m$, and $p_{\pm}=\frac{1}{\sqrt{2}}(p_{x}\pm ip_{y})$.
		
\section{Matrix elements of $\gamma q\to q'$ for proton electric form factor}\label{app:T form factor}
The matrix elements of the electromagnetic transition $\gamma q\to q'$ for the proton electric form factor are derived in detail,
\begin{alignat}{5}\label{eq:T}
	&T^B_{\uparrow\uparrow}&&=&&\ \left[ \dfrac{(E'+m)(E+m)}{4E'E}\right]^\frac{1}{2}\left( 1+ \dfrac{p_z'p_z+2p_-'p_+}{(E'+m)(E+m)}\right), \nonumber\\
	&T^B_{\uparrow\downarrow}&&=&&\ \left[ \dfrac{(E'+m)(E+m)}{4E'E}\right]^\frac{1}{2}\left(  \dfrac{\sqrt{2}(p_z'p_--p_-'p_z)}{(E'+m)(E+m)}\right), \nonumber\\
	&T^B_{\downarrow\uparrow}&&=&&\ \left[ \dfrac{(E'+m)(E+m)}{4E'E}\right]^\frac{1}{2}\left(  \dfrac{\sqrt{2}(-p_z'p_++p_+'p_z)}{(E'+m)(E+m)}\right), \nonumber\\
	&T^B_{\downarrow\downarrow}&&=&&\ \left[ \dfrac{(E'+m)(E+m)}{4E'E}\right]^\frac{1}{2}\left( 1+\dfrac{p_z'p_z+2p_+'p_-}{(E'+m)(E+m)}\right),
\end{alignat}
where $E$ and $E'$ respectively are the energies of the initial and final interacting quarks with the dynamical quark mass of $u$ and $d$ quarks as $m$.
The momentum $p_{\pm}$ are defined by $p_{\pm}=\frac{1}{\sqrt{2}}(p_{x}\pm ip_{y})$.
		
\end{appendix}
	
\bibliographystyle{unsrtnat}
\bibliography{bibtex}

\begin{thebibliography}{28}
\providecommand{\natexlab}[1]{#1}
\providecommand{\url}[1]{\texttt{#1}}
\expandafter\ifx\csname urlstyle\endcsname\relax
  \providecommand{\doi}[1]{doi: #1}\else
  \providecommand{\doi}{doi: \begingroup \urlstyle{rm}\Url}\fi

\bibitem[Zyla et~al.(2020)]{Zyla:2020zbs}
P.A. Zyla et~al.
\newblock {Review of Particle Physics}.
\newblock \emph{PTEP}, 2020\penalty0 (8):\penalty0 083C01, 2020.
\newblock \doi{10.1093/ptep/ptaa104}.

\bibitem[Aznauryan et~al.(2009)Aznauryan, Burkert, Biselli, Egiyan, Joo, Kim,
  Park, Smith, Ungaro, Adhikari, Anghinolfi, Avakian, Ball, Battaglieri,
  Batourine, et~al.]{PhysRevC.80.055203}
I.~G. Aznauryan, V.~D. Burkert, A.~S. Biselli, H.~Egiyan, K.~Joo, W.~Kim,
  K.~Park, L.~C. Smith, M.~Ungaro, K.~P. Adhikari, M.~Anghinolfi, H.~Avakian,
  J.~Ball, M.~Battaglieri, V.~Batourine, et~al.
\newblock Electroexcitation of nucleon resonances from clas data on single pion
  electroproduction.
\newblock \emph{Phys. Rev. C}, 80:\penalty0 055203, Nov 2009.
\newblock \doi{10.1103/PhysRevC.80.055203}.
\newblock URL \url{https://link.aps.org/doi/10.1103/PhysRevC.80.055203}.

\bibitem[Mokeev et~al.(2012)Mokeev, Burkert, Elouadrhiri, Fedotov, Golovatch,
  Gothe, Ishkhanov, Isupov, Adhikari, Aghasyan, Anghinolfi, Avakian,
  Baghdasaryan, Ball, Baltzell, et~al.]{PhysRevC.86.035203}
V.~I. Mokeev, V.~D. Burkert, L.~Elouadrhiri, G.~V. Fedotov, E.~N. Golovatch,
  R.~W. Gothe, B.~S. Ishkhanov, E.~L. Isupov, K.~P. Adhikari, M.~Aghasyan,
  M.~Anghinolfi, H.~Avakian, H.~Baghdasaryan, J.~Ball, N.~A. Baltzell, et~al.
\newblock Experimental study of the ${P}_{11}(1440)$ and ${D}_{13}(1520)$
  resonances from the clas data on
  $ep\ensuremath{\rightarrow}{e}^{\ensuremath{'}}{\ensuremath{\pi}}^{+}{\ensuremath{\pi}}^{\ensuremath{-}}{p}^{\ensuremath{'}}$.
\newblock \emph{Phys. Rev. C}, 86:\penalty0 035203, Sep 2012.
\newblock \doi{10.1103/PhysRevC.86.035203}.
\newblock URL \url{https://link.aps.org/doi/10.1103/PhysRevC.86.035203}.

\bibitem[Mokeev et~al.(2016)Mokeev, Burkert, Carman, Elouadrhiri, Fedotov,
  Golovatch, Gothe, Hicks, Ishkhanov, Isupov, and
  Skorodumina]{PhysRevC.93.025206}
V.~I. Mokeev, V.~D. Burkert, D.~S. Carman, L.~Elouadrhiri, G.~V. Fedotov, E.~N.
  Golovatch, R.~W. Gothe, K.~Hicks, B.~S. Ishkhanov, E.~L. Isupov, and Iu.
  Skorodumina.
\newblock New results from the studies of the $n(1440)1/{2}^{+},
  n(1520)3/{2}^{\ensuremath{-}}$, and
  $\mathrm{\ensuremath{\Delta}}(1620)1/{2}^{\ensuremath{-}}$ resonances in
  exclusive
  $ep\ensuremath{\rightarrow}{e}^{\ensuremath{'}}{p}^{\ensuremath{'}}{\ensuremath{\pi}}^{+}{\ensuremath{\pi}}^{\ensuremath{-}}$
  electroproduction with the clas detector.
\newblock \emph{Phys. Rev. C}, 93:\penalty0 025206, Feb 2016.
\newblock \doi{10.1103/PhysRevC.93.025206}.
\newblock URL \url{https://link.aps.org/doi/10.1103/PhysRevC.93.025206}.

\bibitem[Tiator et~al.(2011)Tiator, Drechsel, Kamalov, and
  Vanderhaeghen]{Tiator:2011pw}
L.~Tiator, D.~Drechsel, S.~S. Kamalov, and M.~Vanderhaeghen.
\newblock {Electromagnetic Excitation of Nucleon Resonances}.
\newblock \emph{Eur. Phys. J. ST}, 198:\penalty0 141--170, 2011.
\newblock \doi{10.1140/epjst/e2011-01488-9}.

\bibitem[Tiator et~al.(2009)Tiator, Drechsel, Kamalov, and
  Vanderhaeghen]{Tiator:2009mt}
L.~Tiator, D.~Drechsel, S.~S. Kamalov, and M.~Vanderhaeghen.
\newblock {Baryon Resonance Analysis from MAID}.
\newblock \emph{Chin. Phys. C}, 33:\penalty0 1069--1076, 2009.
\newblock \doi{10.1088/1674-1137/33/12/005}.

\bibitem[Aznauryan(2007)]{PhysRevC.76.025212}
I.~G. Aznauryan.
\newblock Electroexcitation of the roper resonance in relativistic quark
  models.
\newblock \emph{Phys. Rev. C}, 76:\penalty0 025212, Aug 2007.
\newblock \doi{10.1103/PhysRevC.76.025212}.
\newblock URL \url{https://link.aps.org/doi/10.1103/PhysRevC.76.025212}.

\bibitem[Aznauryan et~al.(2008)Aznauryan, Burkert, Kim, Park, Adams, Amaryan,
  Ambrozewicz, Anghinolfi, Asryan, Avakian, Bagdasaryan, Baillie, Ball,
  Baltzell, Barrow, et~al.]{PhysRevC.78.045209}
I.~G. Aznauryan, V.~D. Burkert, W.~Kim, K.~Park, G.~Adams, M.~J. Amaryan,
  P.~Ambrozewicz, M.~Anghinolfi, G.~Asryan, H.~Avakian, H.~Bagdasaryan,
  N.~Baillie, J.~P. Ball, N.~A. Baltzell, S.~Barrow, et~al.
\newblock Electroexcitation of the roper resonance for $1.7 < {Q}^{2} <4.5$
  {GeV}$^{2}$ in $ep\to en\pi^+$.
\newblock \emph{Phys. Rev. C}, 78:\penalty0 045209, Oct 2008.
\newblock \doi{10.1103/PhysRevC.78.045209}.
\newblock URL \url{https://link.aps.org/doi/10.1103/PhysRevC.78.045209}.

\bibitem[Capstick et~al.(2007)Capstick, Keister, and Morel]{Capstick_2007}
Simon Capstick, B~D Keister, and Danielle Morel.
\newblock Nucleon to resonance form factor calculations.
\newblock \emph{Journal of Physics: Conference Series}, 69:\penalty0 012016,
  may 2007.
\newblock \doi{10.1088/1742-6596/69/1/012016}.
\newblock URL \url{https://doi.org/10.1088/1742-6596/69/1/012016}.

\bibitem[Ramalho and Tsushima(2010)]{PhysRevD.81.074020}
G.~Ramalho and K.~Tsushima.
\newblock Valence quark contributions for the
  $\ensuremath{\gamma}n\ensuremath{\rightarrow}{P}_{11}(1440)$ form factors.
\newblock \emph{Phys. Rev. D}, 81:\penalty0 074020, Apr 2010.
\newblock \doi{10.1103/PhysRevD.81.074020}.
\newblock URL \url{https://link.aps.org/doi/10.1103/PhysRevD.81.074020}.

\bibitem[Golli et~al.(2009)Golli, Sirca, and Fiolhais]{Golli:2009uk}
B.~Golli, S.~Sirca, and M.~Fiolhais.
\newblock {Pion electro-production in the Roper region in chiral quark models}.
\newblock \emph{Eur. Phys. J. A}, 42:\penalty0 185--193, 2009.
\newblock \doi{10.1140/epja/i2009-10878-2}.

\bibitem[Burkert and Roberts(2019)]{RevModPhys.91.011003}
Volker~D. Burkert and Craig~D. Roberts.
\newblock Colloquium: Roper resonance: Toward a solution to the fifty year
  puzzle.
\newblock \emph{Rev. Mod. Phys.}, 91:\penalty0 011003, Mar 2019.
\newblock \doi{10.1103/RevModPhys.91.011003}.
\newblock URL \url{https://link.aps.org/doi/10.1103/RevModPhys.91.011003}.

\bibitem[Li and Riska(2006)]{PhysRevC.74.015202}
Q.~B. Li and D.~O. Riska.
\newblock Role of $q\overline{q}$ components in the $n(1440)$ resonance.
\newblock \emph{Phys. Rev. C}, 74:\penalty0 015202, Jul 2006.
\newblock \doi{10.1103/PhysRevC.74.015202}.
\newblock URL \url{https://link.aps.org/doi/10.1103/PhysRevC.74.015202}.

\bibitem[Obukhovsky et~al.(2011)Obukhovsky, Faessler, Fedorov, Gutsche, and
  Lyubovitskij]{PhysRevD.84.014004}
Igor~T. Obukhovsky, Amand Faessler, Dimitry~K. Fedorov, Thomas Gutsche, and
  Valery~E. Lyubovitskij.
\newblock Electroproduction of the roper resonance on the proton: The role of
  the three-quark core and the molecular $n\ensuremath{\sigma}$ component.
\newblock \emph{Phys. Rev. D}, 84:\penalty0 014004, Jul 2011.
\newblock \doi{10.1103/PhysRevD.84.014004}.
\newblock URL \url{https://link.aps.org/doi/10.1103/PhysRevD.84.014004}.

\bibitem[Cano and Gonzalez(1998)]{Cano:1998wz}
F.~Cano and P.~Gonzalez.
\newblock {A Consistent explanation of the Roper phenomenology}.
\newblock \emph{Phys. Lett. B}, 431:\penalty0 270--276, 1998.
\newblock \doi{10.1016/S0370-2693(98)00574-7}.

\bibitem[\ifmmode~\check{S}\else \v{S}\fi{}tajner
  et~al.(2017)\ifmmode~\check{S}\else \v{S}\fi{}tajner, Achenbach, Beranek,
  Beri\ifmmode \check{c}\else \v{c}\fi{}i\ifmmode~\check{c}\else \v{c}\fi{},
  Bernauer, Bosnar, B\"ohm, Correa, Denig, Distler, Esser, Fonvieille,
  Friedrich, Fri\ifmmode \check{s}\else \v{s}\fi{}\ifmmode \check{c}\else
  \v{c}\fi{}i\ifmmode~\acute{c}\else \'{c}\fi{}, Kegel, Kohl, Merkel, ,
  et~al.]{PhysRevLett.119.022001}
S.~\ifmmode~\check{S}\else \v{S}\fi{}tajner, P.~Achenbach, T.~Beranek,
  J.~Beri\ifmmode \check{c}\else \v{c}\fi{}i\ifmmode~\check{c}\else \v{c}\fi{},
  J.~C. Bernauer, D.~Bosnar, R.~B\"ohm, L.~Correa, A.~Denig, M.~O. Distler,
  A.~Esser, H.~Fonvieille, J.~M. Friedrich, I.~Fri\ifmmode \check{s}\else
  \v{s}\fi{}\ifmmode \check{c}\else \v{c}\fi{}i\ifmmode~\acute{c}\else
  \'{c}\fi{}, S.~Kegel, Y.~Kohl, H.~Merkel, , et~al.
\newblock Beam-recoil polarization measurement of ${\ensuremath{\pi}}^{0}$
  electroproduction on the proton in the region of the roper resonance.
\newblock \emph{Phys. Rev. Lett.}, 119:\penalty0 022001, Jul 2017.
\newblock \doi{10.1103/PhysRevLett.119.022001}.
\newblock URL \url{https://link.aps.org/doi/10.1103/PhysRevLett.119.022001}.

\bibitem[Sarantsev et~al.(2008)]{Sarantsev:2007aa}
A.~V. Sarantsev et~al.
\newblock {New results on the Roper resonance and the P(11) partial wave}.
\newblock \emph{Phys. Lett. B}, 659:\penalty0 94--100, 2008.
\newblock \doi{10.1016/j.physletb.2007.11.055}.

\bibitem[Capstick and Roberts(2000)]{Capstick:2000qj}
Simon Capstick and W.~Roberts.
\newblock {Quark models of baryon masses and decays}.
\newblock \emph{Prog. Part. Nucl. Phys.}, 45:\penalty0 S241--S331, 2000.
\newblock \doi{10.1016/S0146-6410(00)00109-5}.

\bibitem[Xu et~al.(2019)Xu, Kaewsnod, Liu, Srisuphaphon, Limphirat, and
  Yan]{PhysRevC.100.065207}
K.~Xu, A.~Kaewsnod, X.~Y. Liu, S.~Srisuphaphon, A.~Limphirat, and Y.~Yan.
\newblock Complete basis for the pentaquark wave function in a group theory
  approach.
\newblock \emph{Phys. Rev. C}, 100:\penalty0 065207, Dec 2019.
\newblock \doi{10.1103/PhysRevC.100.065207}.
\newblock URL \url{https://link.aps.org/doi/10.1103/PhysRevC.100.065207}.

\bibitem[Xu et~al.(2020)Xu, Kaewsnod, Zhao, Liu, Srisuphaphon, Limphirat, and
  Yan]{PhysRevD.101.076025}
K.~Xu, A.~Kaewsnod, Z.~Zhao, X.~Y. Liu, S.~Srisuphaphon, A.~Limphirat, and
  Y.~Yan.
\newblock Pentaquark components in low-lying baryon resonances.
\newblock \emph{Phys. Rev. D}, 101:\penalty0 076025, Apr 2020.
\newblock \doi{10.1103/PhysRevD.101.076025}.
\newblock URL \url{https://link.aps.org/doi/10.1103/PhysRevD.101.076025}.

\bibitem[Janssens et~al.(1966)Janssens, Hofstadter, Hughes, and
  Yearian]{PhysRev.142.922}
T.~Janssens, R.~Hofstadter, E.~B. Hughes, and M.~R. Yearian.
\newblock Proton form factors from elastic electron-proton scattering.
\newblock \emph{Phys. Rev.}, 142:\penalty0 922--931, Feb 1966.
\newblock \doi{10.1103/PhysRev.142.922}.
\newblock URL \url{https://link.aps.org/doi/10.1103/PhysRev.142.922}.

\bibitem[Berger et~al.(1971)Berger, Burkert, Knop, Langenbeck, and
  Rith]{Berger:1971kr}
C~Berger, V.~Burkert, G.~Knop, B.~Langenbeck, and K.~Rith.
\newblock {Electromagnetic form-factors of the proton at squared four momentum
  transfers between 10-fm**-2 and 50 fm**-2}.
\newblock \emph{Phys. Lett. B}, 35:\penalty0 87--89, 1971.
\newblock \doi{10.1016/0370-2693(71)90448-5}.

\bibitem[Price et~al.(1971)Price, Dunning, Goitein, Hanson, Kirk, and
  Wilson]{Price:1971zk}
L.~E. Price, J.~R. Dunning, M.~Goitein, K.~Hanson, T.~Kirk, and R.~Wilson.
\newblock {Backward-angle electron-proton elastic scattering and proton
  electromagnetic form-factors}.
\newblock \emph{Phys. Rev. D}, 4:\penalty0 45--53, 1971.
\newblock \doi{10.1103/PhysRevD.4.45}.

\bibitem[Hanson et~al.(1973)Hanson, Dunning, Goitein, Kirk, Price, and
  Wilson]{PhysRevD.8.753}
K.~M. Hanson, J.~R. Dunning, M.~Goitein, T.~Kirk, L.~E. Price, and Richard
  Wilson.
\newblock Large-angle quasielastic electron-deuteron scattering.
\newblock \emph{Phys. Rev. D}, 8:\penalty0 753--778, Aug 1973.
\newblock \doi{10.1103/PhysRevD.8.753}.
\newblock URL \url{https://link.aps.org/doi/10.1103/PhysRevD.8.753}.

\bibitem[Murphy et~al.(1974)Murphy, Shin, and Skopik]{PhysRevC.9.2125}
J.~J. Murphy, Y.~M. Shin, and D.~M. Skopik.
\newblock Proton form factor from 0.15 to 0.79
  ${\mathrm{fm}}^{\ensuremath{-}2}$.
\newblock \emph{Phys. Rev. C}, 9:\penalty0 2125--2129, Jun 1974.
\newblock \doi{10.1103/PhysRevC.9.2125}.
\newblock URL \url{https://link.aps.org/doi/10.1103/PhysRevC.9.2125}.

\bibitem[Höhler et~al.(1976)Höhler, Pietarinen, Sabba-Stefanescu, Borkowski,
  Simon, Walther, and Wendling]{HOHLER1976505}
G.~Höhler, E.~Pietarinen, I.~Sabba-Stefanescu, F.~Borkowski, G.G. Simon, V.H.
  Walther, and R.D. Wendling.
\newblock Analysis of electromagnetic nucleon form factors.
\newblock \emph{Nuclear Physics B}, 114\penalty0 (3):\penalty0 505--534, 1976.
\newblock ISSN 0550-3213.
\newblock \doi{https://doi.org/10.1016/0550-3213(76)90449-1}.
\newblock URL
  \url{https://www.sciencedirect.com/science/article/pii/0550321376904491}.

\bibitem[Simon et~al.(1980)Simon, Schmitt, Borkowski, and
  Walther]{Simon:1980hu}
G.~G. Simon, C.~Schmitt, F.~Borkowski, and V.~H. Walther.
\newblock {Absolute electron Proton Cross-Sections at Low Momentum Transfer
  Measured with a High Pressure Gas Target System}.
\newblock \emph{Nucl. Phys. A}, 333:\penalty0 381--391, 1980.
\newblock \doi{10.1016/0375-9474(80)90104-9}.

\bibitem[Walker et~al.(1994)Walker, Filippone, Jourdan, Milner, McKeown,
  Potterveld, Andivahis, Arnold, Benton, Bosted, deChambrier, Lung, Rock,
  Szalata, Para, et~al.]{PhysRevD.49.5671}
R.~C. Walker, B.~W. Filippone, J.~Jourdan, R.~Milner, R.~McKeown,
  D.~Potterveld, L.~Andivahis, R.~Arnold, D.~Benton, P.~Bosted, G.~deChambrier,
  A.~Lung, S.~E. Rock, Z.~M. Szalata, A.~Para, et~al.
\newblock Measurements of the proton elastic form factors for
  $1\ensuremath{\le}{Q}^{2}\ensuremath{\le}3$
  ${(\mathrm{G}\mathrm{e}\mathrm{V}/\mathit{c})}^{2}$ at slac.
\newblock \emph{Phys. Rev. D}, 49:\penalty0 5671--5689, Jun 1994.
\newblock \doi{10.1103/PhysRevD.49.5671}.
\newblock URL \url{https://link.aps.org/doi/10.1103/PhysRevD.49.5671}.

\end{thebibliography}
	
\end{document}